# Theoretical Estimates of Integrated Sachs-Wolfe Effect Detection through EMU-ASKAP Survey with Confusion, Position Uncertainty, Shot Noise and SNR analysis.

**SYED FAISAL UR RAHMAN**

**6/19/2014**

Institution Address: University Road, University of Karachi, Institute of Space and Planetary Astrophysics (ISPA), 75270.

Author Address: B-58, C.G.E.C.H.S, Block 10-A, Gulshan e Iqbal, Karachi, 75300.

Email: faisalrahman36@hotmail.com

Phone: +92-21-34814531

Cell Number: +92-335-2107453

# Theoretical Estimates of Integrated Sachs-Wolfe Effect Detection through EMU-ASKAP Survey with Confusion, Position Uncertainty, Shot Noise and SNR analysis.

Syed Faisal ur Rahman

Institute of Space and Planetary Astrophysics, University of Karachi
faisalrahman36@hotmail.com

## ABSTRACT

Detection of the late time integrated Sachs-Wolfe effect (ISW) is an active area of study related to Large Scale Structures (LSS).The ISW effect can be studied by observing the non-zero cross-correlation between cosmic microwave background (CMB) anisotropies with tracers of mass field, such as galaxy survey data. We plan to study this effect by cross-correlating the CMB data and related cosmological parameters as delineated by the Wilkinson microwave anisotropy probe (WMAP) with the upcoming evolutionary map of the universe (EMU) survey planned for Australian square kilometer array Pathfinder(ASKAP). EMU-ASKAP is planned to conduct a deep radio continuum survey with rms 10μJy/beam. The survey is planned to cover the entire southern sky, extending to North + 30 degree declination. To infer the expected redshift distribution (dN/dz) and differential source count (S) of the survey that can be extracted from the EMU galaxies, we use data from S-cubed simulation of extra-galactic radio continuum ($S^3$- SEX) for square kilometer array design studies (SKADS). We also calculate various parameters including galaxy survey shot noise, rms confusion uncertainty, and position uncertainty for the survey which can help in understanding the accuracy and analyzing the results of the data. We also discuss Signal to noise ratios over range of maximum redshifts and maximum multipole values with some discussion on constraints over dark energy density parameter (ΩΛ) and matter density parameter (Ωb).

## 1. INTRODUCTION

One of the biggest goals of modern cosmology is the understanding of dark energy and the distribution of matter in the universe. Galaxy surveys provide some of the most important parameters, especially in combination with cosmic microwave background surveys, which can be used to calculate phenomena such as Integrated Sachs Wolfe effect [38], gravitational lensing, galaxy matter power spectrum and others. These phenomena help in measuring parameters related to dark energy. For most such analyses, it is important to know about the galaxy redshift distribution and how galaxy redshift bias can affect the analysis. As first presented by [12], a suitable way of measuring linear ISW effect is through the cross-correlation of CMB data with the tracers of large scale structures (LSS). Some detailed studies have been made to identify benchmarks for assessing the qualities of LSS surveys and their ability to measure ISW effect like[1], provided a detailed theoretical framework for observing ISW effect, [13] and[26] investigated power of some future surveys related to ISW detection. A study by Douspis provided 4σ estimates for ideal surveys with median redshift > 0.84 and fsky> 0.35[13]. Earlier studies of ISW effect,(for e.g. [17],[31],[39],[5] and[4]) showed average 3-3.5 σ detections and some recent ones like [18], [23],[20] and[15],provided much improved 4.5-5 σ average.

However, there is a gap in the literature to see the effects of shot noise on the significance of observational results with variable maximum redshift and multipole ranges especially using non-ideal (non-uniform redshift distribution, finite source count) conditions of an upcoming survey. In this study we focus on EMU-ASKAP parameters and discussion will be related to observational issues like shot noise, confusion and position accuracy, and their relation to theoretical estimates. We will especially see how Signal to Noise ratios will be affected by shot noise which will demonstrate the benefits of deeper surveys like EMU-ASKAP. Often theoretical papers neglect observational constraints and in this paper, we have tried to bridge this gap. We think this study will be helpful for both observational astronomers in critically analyzing their own results. Also theorists will realize the importance of considering limitations related to observational issues which will help them in improving their approximations.

## 2. EMU-ASKAP SURVEY AND DATA FOR ESTIMATES

Australian Square Kilometer Array Pathfinder (ASKAP). Evolutionary Map of the Universe Survey (EMU) will cover around 75% (3π steradian) of the sky. The survey is planned to cover the entire southern sky, extending to North + 30 degree declination. The survey covers roughly the same area but will be around 45 times deeper than previously conducted NRAO VLA sky survey (NVSS) [10],[32],[11]. ASKAP has a total of 36 antennas with an individual diameter of 12 m and 113 sq. m dish area [14].The effective area for ASKAP's full 36 antenna resolution with an aperture efficiency of 0.8 is 3211 sq. m. Maximum baseline for full 36 antenna configuration is 6 km. When operational, EMU-ASKAP, with ≈ 10μJy/beam sensitivity, will provide the most sensitive radio galaxy surveys of its time before the Square Kilometer Array (SKA. EMU-ASKAP is one of the two top ranked scientific goals of ASKAP along with WALLABY [14]. Here we use $S^3$ -SEX database [40] for SKADS to extract semi-empirical galaxy redshift distribution of given 400 square degrees of sky and then we extrapolate it to give per steradian per redshift distribution to be used in the measurement of theoretical linear ISW effect.

We use cosmological parameters as delineated by 9 years WMAP survey data [21] to calculate theoretical angular cross- correlation between CMB and EMU-ASKAP galaxy survey data. In following sections we will use wmap9+bao+h0+spt+act+snls3 parameters with ΛCDM cosmology. PLANCK [33] has also announced its initial results, but more detailed probe data and analysis will be released in the future so we focused more on WMAP9. The paper is more about EMU-ASKAP's ability to measure the effect and focuses on some technical issues relevant to observations and error analysis.

# 3. CONFUSION AND POSITION ACCURACY

Sensitivity of a survey can help in analyzing depth and integration time, but to clearly resolve observed sources we need to keep accounts for confusion parameter. A high sensitivity survey with low resolution can increase position uncertainty due to increase in confusion. To calculate rms confusion, we need to fit power law curve for differential source count of the survey. We need to measure differential source count in $Jy^{-1}\ Sr^{-1}$as:

$$n(s) = \frac{dN}{dS} = kS^{-\gamma} \tag{1}$$

We first use SKADS differential source counts with flux density range$-4.3 < \log(S) < -3.4$and obtain k=57.24 and $\gamma$=2.18. [9] Uses flux range of$-6 < \log(S) < -4$withk=1000 and $\gamma$ =1.9. Similarly, [25] obtained k=8.23 and $\gamma$ =2.4. In another study [30] measured k=57 and $\gamma$=2.2.

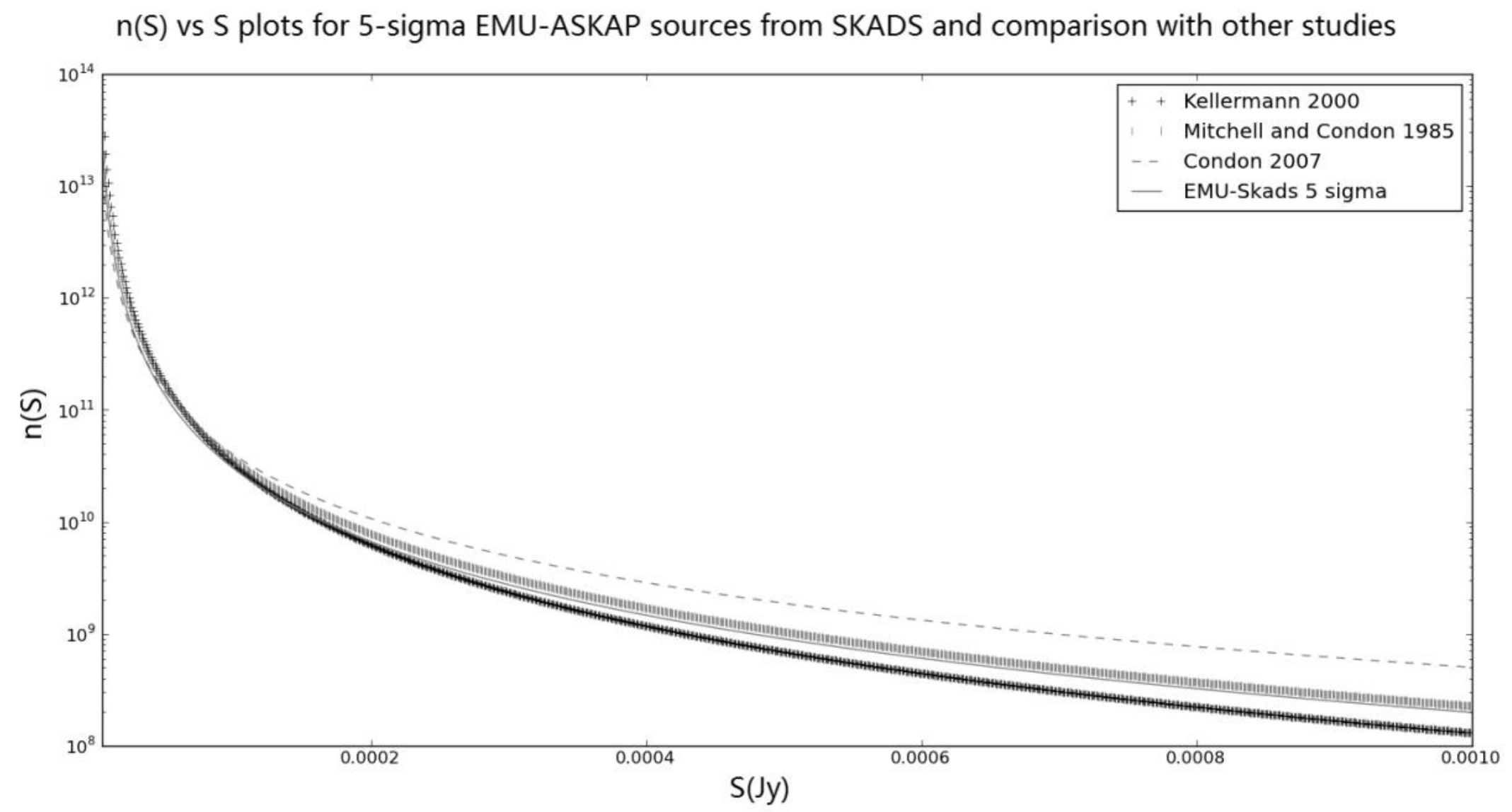

Figure 1-logarithmic differential source count plot for EMU-SKADS 5 sigma and comparison with previous studies. Here we used ΔS= .000005 Jy.

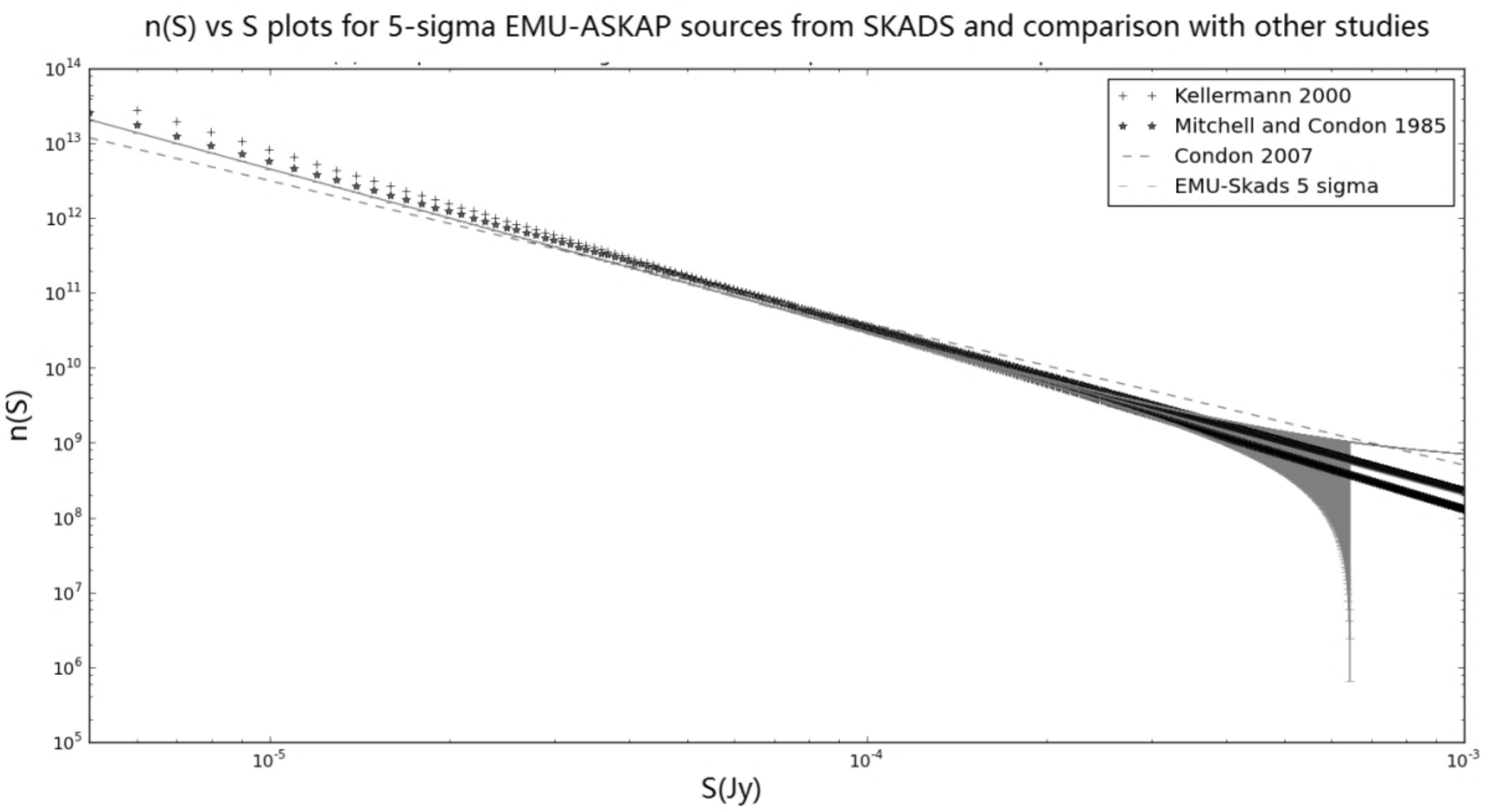

Figure 2-log-log plot of differential source count with standard deviation for EMU-SKADS 5 sigma.

The fluctuations related to observed sky-brightness caused by the presence of faint sources in the telescope beam are called confusion. These fluctuations play a vital role in densely populated areas like galactic plane or surroundings. The power law distribution provides a useful method to estimate rms confusion. Using the k and γ values we can calculate rms confusion [7]:

$$\sigma c = \left(\frac{q^{3-\gamma}}{3-\gamma}\right)^{\frac{1}{\gamma-1}} (k\Omega e)^{\frac{1}{\gamma-1}} (2)$$

Here q is taken as 5 using Dmax=qσc[7] and Ωe can be calculated by Ωe=Ωb/(γ-1). Ωb can be calculated

using θ = 10 arcsec or 4.85e-05 radians FWHM for EMU-ASKAP

$$\Omega_b = \frac{\pi\theta^2}{4ln2} \qquad (3)$$

In this section, Ωb is for solid angle and should not to be confused with Ωb for the baryon matter density parameter. For Ωb=2.664 e-09 with k and γ as measured in our case, of EMU-SKADS 5 sigma sources, we get Ωe=2.2576e-09 and σc=5.26 μJy

For noise or sensitivity calculation of single beam we have a relation [14]:

$$\sigma n = \sqrt{2}\,\frac{kTsys}{Aeff\sqrt{npol}\Delta\nu t} \quad (4)$$

Where k=1380 Jy $m^2 K^{-1}$ , average $T_{sys}$ is 50K and Δν =300MHz for a continuum survey andfor a dual polarization antenna we have $n_{pol}$=2. $A_{eff}$ can be calculated using the relation:

$$A_{eff} = \eta_a A\sqrt{N(N-1)} \quad (5)$$

Where $\eta_a$ is the aperture efficiency, A=113 sq m is the area of individual antenna for ASKAP and N is the number of antennas used in the configuration. The EMU-ASKAP survey will have a sensitivity of 10μJy/beam [32].

We can obtain total rms σtot as:

$$\sigma^2 tot = \sigma^2 n + \sigma^2 c \quad (6)$$

For σn=10μJy and σc=5.26 μJy, we get σtot≈ 11.3μJy. Position uncertainty σp rms is another useful measure in identification of radio sources. Total rms noise σtot affects σp as:

$$\sigma p \approx \frac{\sigma tot\theta}{2S} \quad (7)$$

For S=56.5μJy or 5σtot and θ=10 arcsec, we get σp=1 arcsec and for S= 50μJy or 5σn we will get rms position uncertainty of σp ≈1.13arcsec. In NVSS survey σp < 1arcsec requires S > 10mJy[10] which shows that with EMU we can go much deeper without compromising on rms position uncertainty.

## 4. SHOT NOISE

Shot noise analysis is required to plot SNR and other ratios while using spherical harmonics and finding Cl values for the cross correlation of CMB- Galaxy and auto correlation of Galaxy data.

$$Shot\ Noise = \frac{\Delta\Omega}{N} \qquad (8)$$

Where, $\Delta\Omega$ =observed area in steradian and N=number of sources observed in $\Delta\Omega$ or N=Ns$\Delta\Omega$. We can also see that shot noise can be written as shot noise=1/Ns where Ns is the number of sources per steradian. Using SKADS database, we obtain Ns = 3121949 for EMU-5σ radio sources with redshift (z) between 0 and 1. Here we get shot noise ≈ 3.20313e-07. Shot noise plays an important role in calculating angular power spectrum covariance especially for large multipole (l) values.

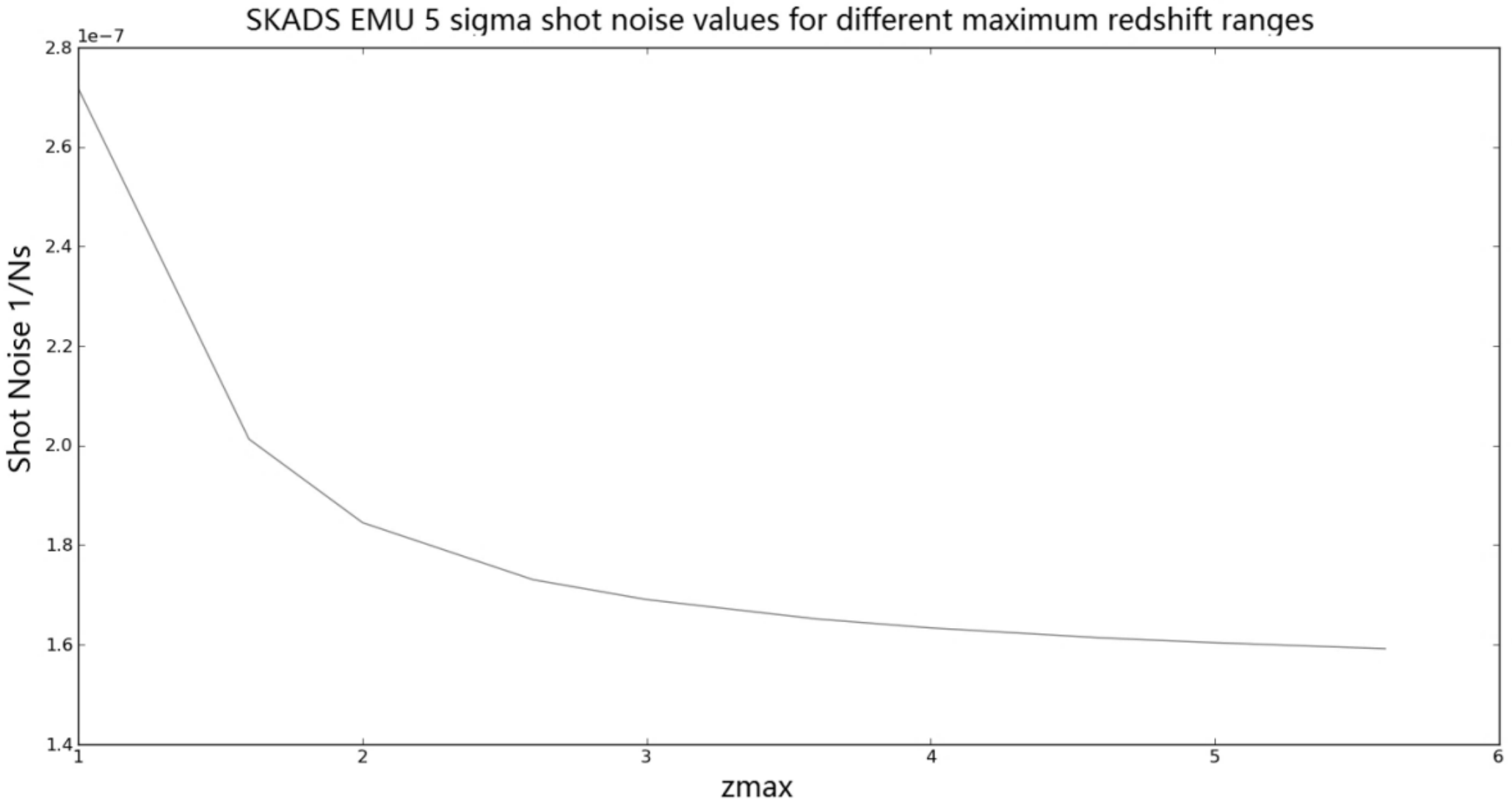

Figure 3-Shot Noise values for different zmax ranges.

It is an obvious thing to note that shot noise decreases with increasing redshift range and increased source count. In an ideal scenario with NS→ ∞ shot noise should be zero or close to zero, but due to constraints like integration time, confusion and position uncertainty, our ability to achieve much higher sensitivity or depth is limited.

## 5. ISW EFFECT ESTIMATES

We can write total temperature perturbation as (for e.g. [16],[37]and for more detailed study read [19]):

$$\frac{\delta T(\boldsymbol{n},\eta 0)}{T} = \frac{1}{4}\delta\gamma(\eta r) + \Phi(\eta r) + \int_{\eta r}^{\eta 0}(\Phi' - \Psi')d\eta + \boldsymbol{nv}\,(\eta r)(9)$$

First two parts of the equation are Sachs-Wolfe effect, third parts is Integrated Sachs-Wolfe effect and the last part is due to the Doppler Effect when baryon-electron-photon medium moves with respect to the conformal Newtonian frame with the velocity v(ηr). Here ηr is the conformal time at decoupling and η0 is current conformal time.

ISW effect explains the blue shifting of photons from the surface of last scattering when they enter gravitational potential wells of large scale structure (LSS). These photons get red shifted when they leave the potential wells. But due to the coherent decay of the gravitational potential well as a result of the accelerating expansion of the universe, photons keep some of the energy which results in their net blue shifting. Reverse happens in the case of gravitational hills or super-voids. This shifting of wavelengths results in CMB temperature anisotropy. Detection of the late time integrated Sachs-Wolfe effect (ISW) is one of the few ways to investigate dark energy and some of its physical properties.

ISW effect can be studied by observing the non- zero cross-correlation between cosmic microwave background (CMB) anisotropies with tracers of mass field, such as galaxy survey data in the form galaxy over/under-density maps. Late Time ISW effect is related to the time when dark energy started to dominate the universe, replacing the domination of matter. Looking at equation 9, we can also see that there will be no ISW effect in matter dominated universe as Φ’ and Ψ’ are 0 in that regime[24]. In other words positive cross-correlation between CMB and LSS data or presence of the ISW effect can be used as a signature of dark energy.

We can calculate cross-correlation power spectrum between WMAP 9 years and EMU-SKADS 5 sigma sources using the relation:

$$Cl^{gt} = 4\pi \int_{kmin}^{kmax} \frac{dk}{k} \Delta^2(k) Wl^g(k) Wl^t(k) \qquad (10)$$

Here is$\Delta^2(k)$the logarithmic matter power spectrum, which can be calculated as:

$$\Delta^2(\mathrm{k}) = \frac{k^3}{2\pi^2} P(k) \qquad (11)$$

Where, P(k) is the matter power spectrum. We use CAMB [27] to obtain P(k). $Wl^g(k)$ and $Wl^t(k)$ represent galaxy and ISW window functions. These window functions can be written as [35]:

$$Wl^g(k) = \int dz \frac{dN}{dz} b(z) g(z) jl(k\chi(\mathrm{z})) \qquad (12)$$

$$Wl^t(k) = 3Tcmb\Omega m \left(\frac{H0}{ck}\right)^2 \int dz G(z) jl(k\chi(\mathrm{z})\,) \qquad (13)$$

Where g(z)=$\frac{D(z)}{D(0)}$,jl(x) is spherical Bessel function, b(z) is galaxy bias and χ(z)=cη(z), with η(z) as conformal loop back time. Also in ISW window function, G(z)=d[D(z)(1+z)/D(0)]/dz Tcmb is the average CMB temperature taken as 2.725 K, c is the speed of light and H0 is the Hubble constant.

Figure 4 shows estimated window functions for ISW signal and EMU-ASKAP 5σ galaxies. We can see the decreasing contribution of window functions with increase in multipole (l) values. This also affects the selection of maximum multipole or lmax range of ISW studies.

We can write η(z) as:

$$\eta(\mathrm{z}) = \int_0^z \frac{dz'}{H(z)} = \int_0^z \frac{dz'}{H0E(z')} \qquad (14)$$

With $E(z)=\sqrt{\Omega_\Lambda+\Omega r(1+z)^4+\Omega m(1+z)^3}$, Ωr is set to zero for our study. For large scales, linear growth factor can be written approximated as [6]:

$$D(z)=\frac{5\Omega m(z)}{2(1+z)}\left\{\Omega m(z)^{\frac{4}{7}}-\Omega\Lambda(z)+\left[1+\frac{\Omega m(z)}{2}\right].\left[1+\frac{\Omega\Lambda(z)}{70}\right]\right\}^{-1} \quad (15)$$

Where,

$\Omega(z) = \Omega m(1+z)^3 / E(z)^2$

$\Omega\Lambda(z) = \Omega\Lambda / E(z)^2$

We use the redshift distribution per steradian (dN/dz) obtained from $S^3$ SEX database of SKADS with $S > 50\mu Jy$ to get semi-empirical distribution for EMU-ASKAP 5σ radio sources. We normalize this function such that $\Sigma N(z)dz = 1$ over the redshift range from 0 to 1. We first estimate $Cl^{gt}$ for redshift range 0 to 1 with monopoles $3 \leq l \leq 100$ and then calculate S/N ratio for this range. Then we use the same formalism to calculate S/N ratios for greater redshift ranges.

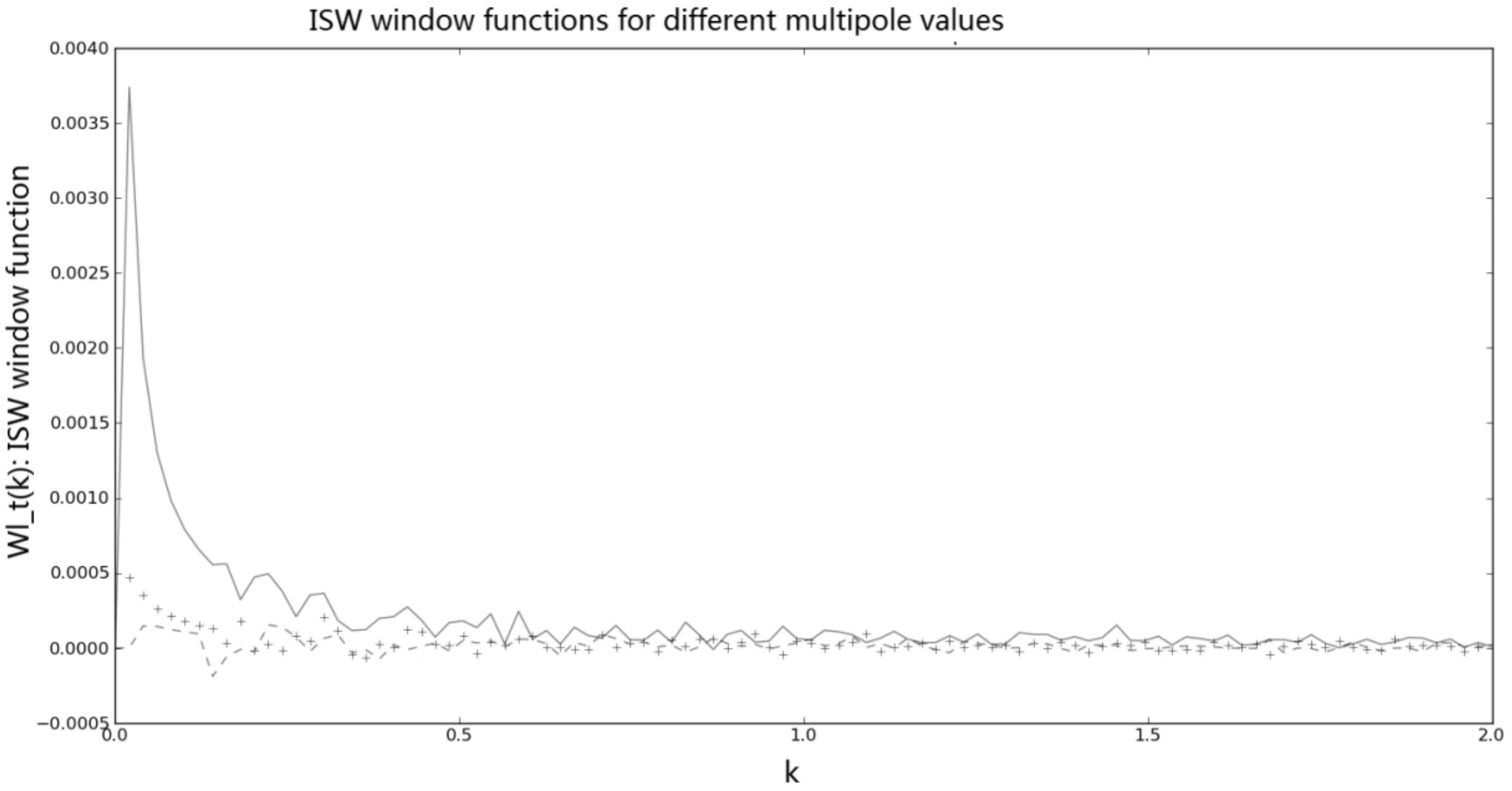

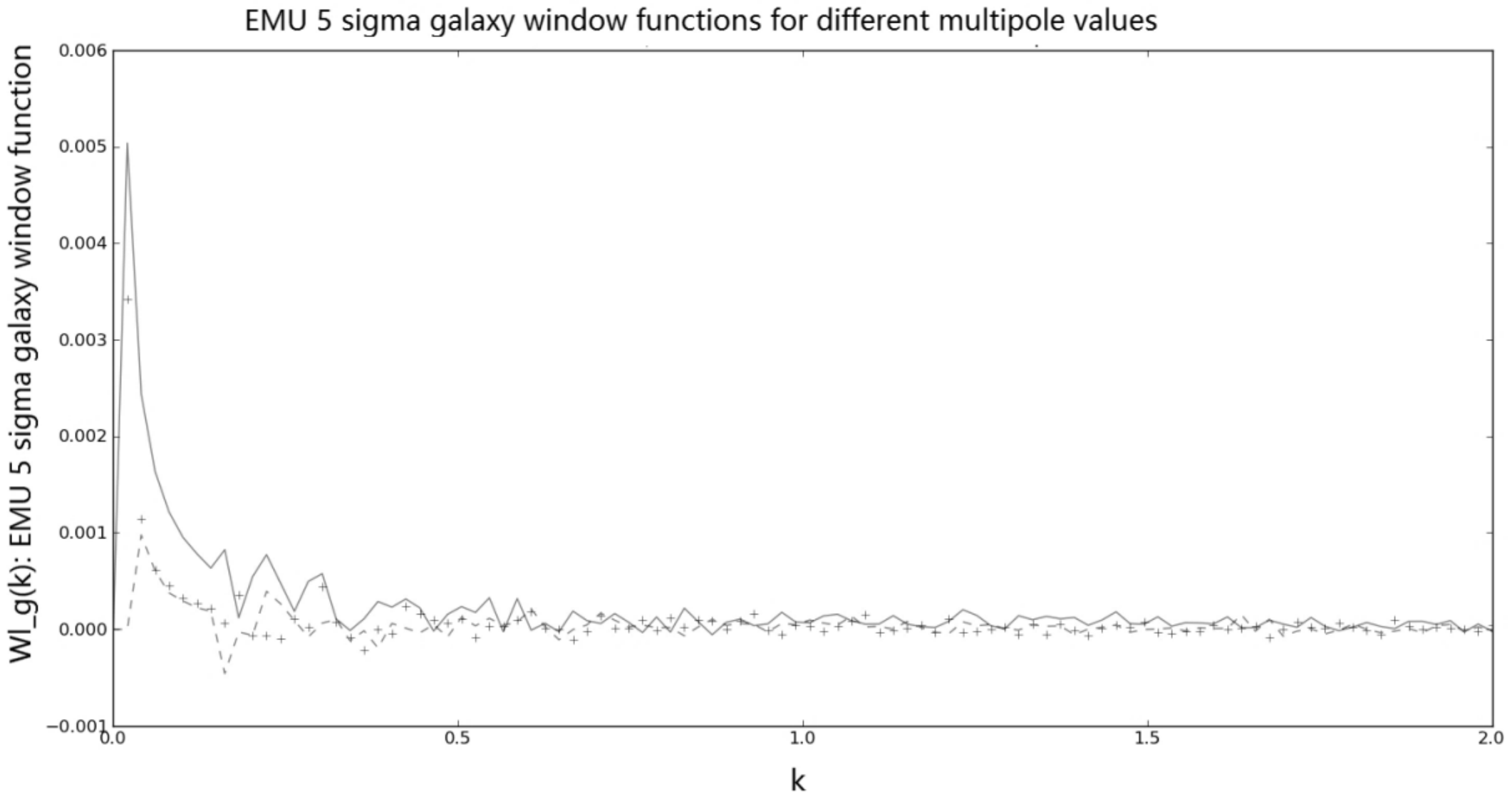

Figure 4-Plots for ISW window function (top) and EMU-SKADS 5 sigma window function (bottom). Solid line represents values for multipole (l)=3, '+ +' represents l=50 and '--' represents l=100.

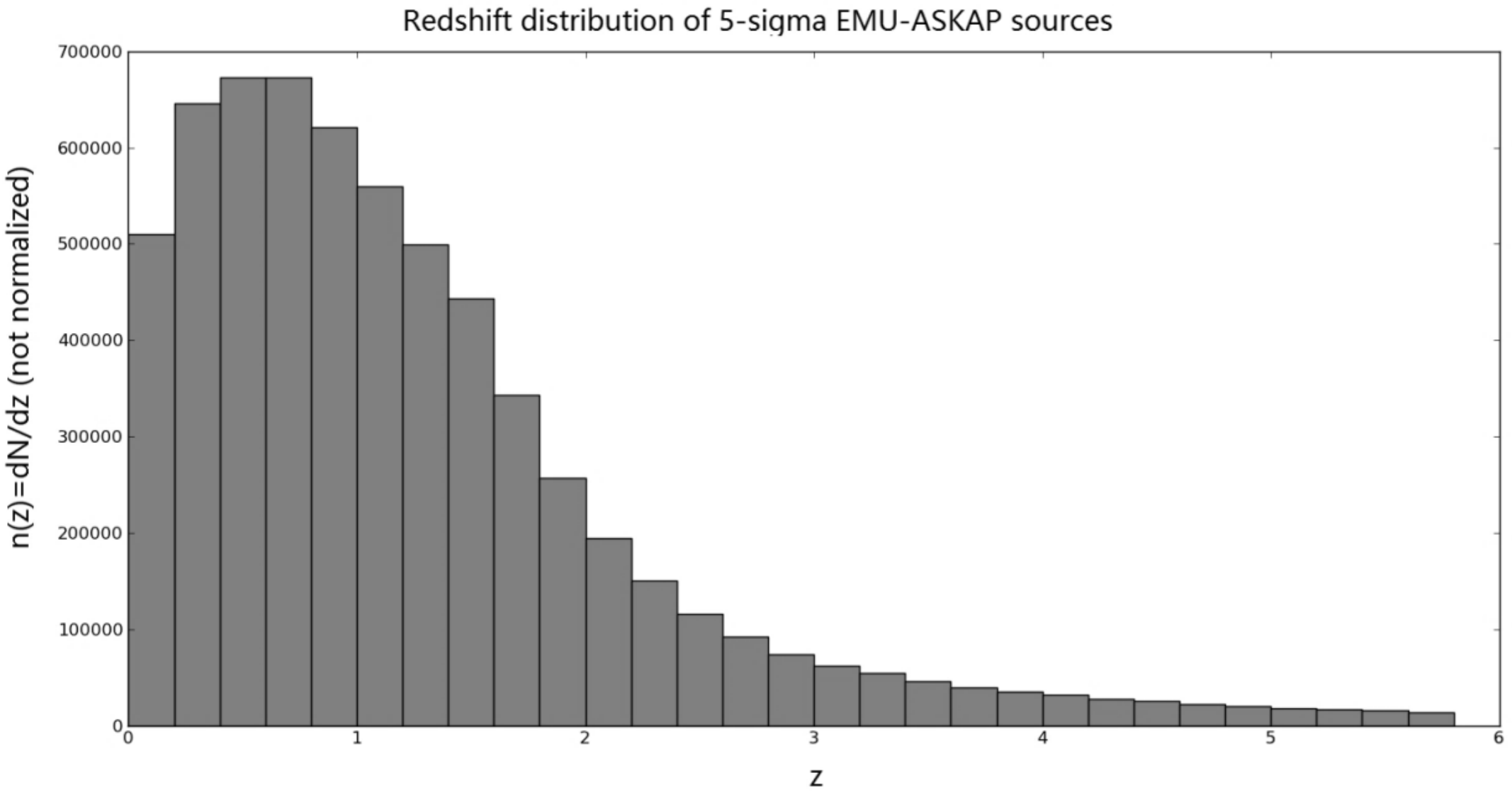

Figure 5-Redshift distribution for EMU- ASKAP 5σ sources obtained from SKADS. Here n(z)=dN/dz with Δz =0.2.

We then use [34] and[40] to obtain bias factor b(z) for EMU 5σ sources. For computational reasons, we use limber approximations of (for e.g. ref [1], [3],[23] and [22]) which give an error of $O(l^{-2})$. The resulting form for $Cl^{gt}$ looks like:

$$Cl^{gt} = \frac{3\Omega m H0^{2} Tcmb}{\left(l+\frac{1}{2}\right)^{2} c^{2}} \int \frac{dz b(z) \boldsymbol{N}(\boldsymbol{z}) H(z)}{c} g(z) G(z) P\left\{\frac{l+\frac{1}{2}}{\chi(z)}\right\} \quad (16)$$

Here k = (l+1/2)/χ from limber approximation (for e.g. [28] and [29])is used.

To obtain cross correlation function CCF(θ ) for $Cl^{gt}$ we use[3]:

$$\mathrm{CCF}(\theta)=\sum_{l} \frac{2l+1}{4\pi} Cl^{gt} Pl(cos\theta) \qquad (17)$$

## 6. ERROR ANALYSIS AND SNR

To obtain error bars and signal to noise ratio (S/N) we use cosmic variance relation [34] and [2]:

$$\Delta Cl^{gt} = \sqrt{\frac{((Cl^{gg}+\frac{1}{NS})Cl^{tt}+(Cl^{gt})^2)}{(2l+1)fsky}} \qquad (18)$$

Here $Cl^{gg}$is galaxy auto correlation and $Cl^{tt}$ is CMB auto correlation calculated using CAMB. Here 1/Ns is shot noise and as discussed earlier, it plays an important role for high 'l' values. We converted $\Delta\, Cl^{gt}$ into θ domain as:

$$\Delta\, Cl^{gt}(\theta)=\sqrt{\sum_l \left(\frac{2l+1}{4\pi}\right)^2 \Delta^2 Cl^{gt} Pl^2(cos\theta)} \qquad (19)$$

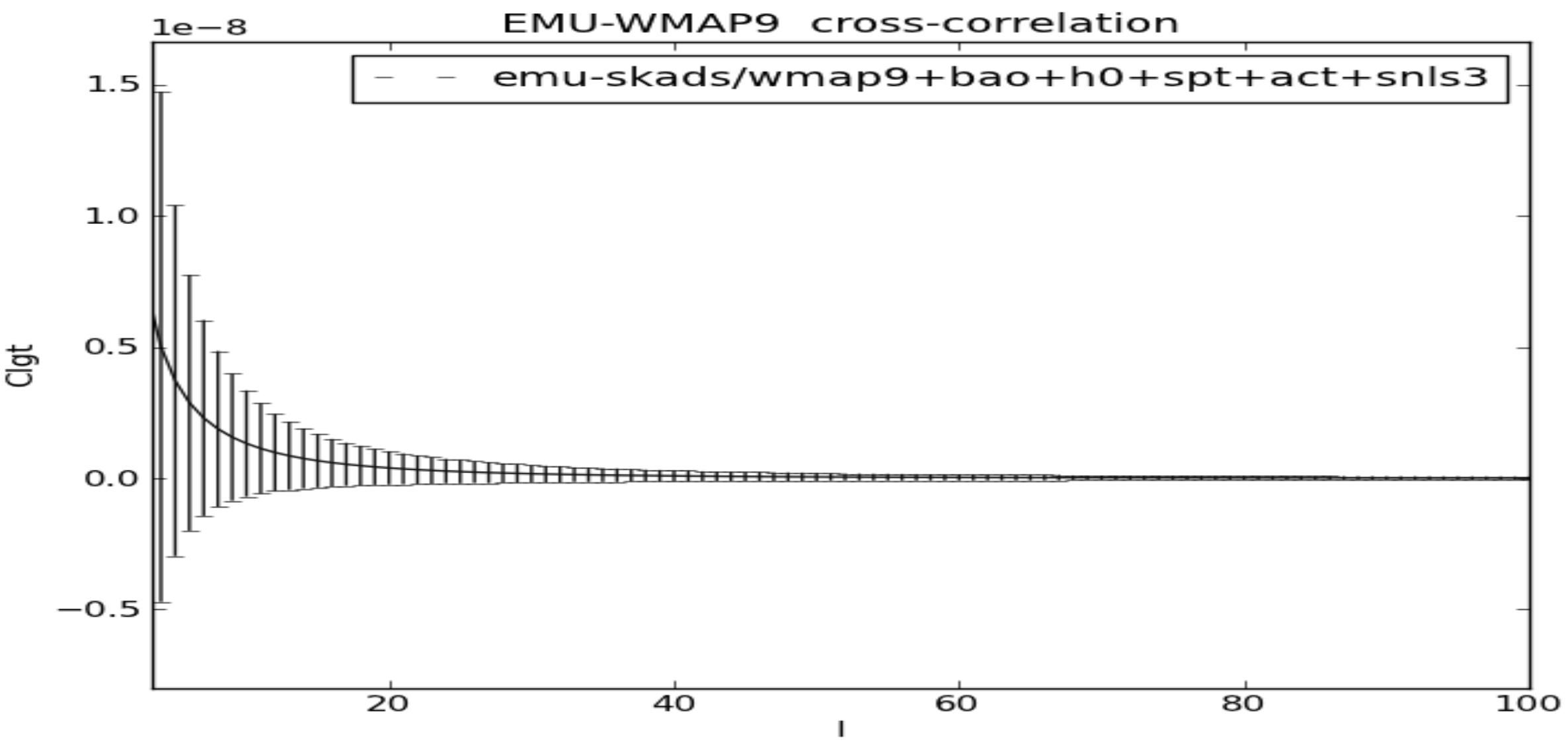

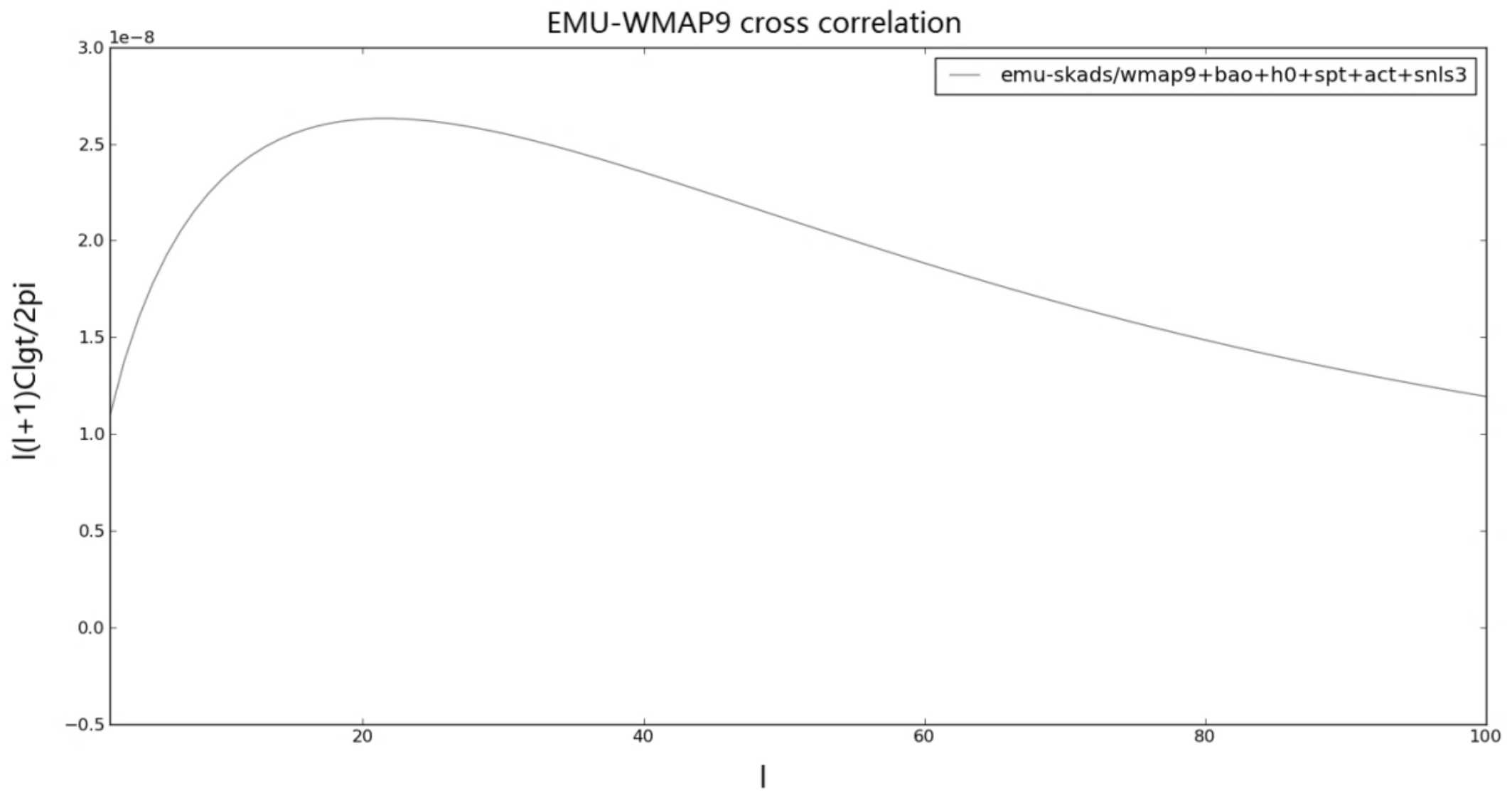

Figure 6-Top: Cross-correlation power spectrum between CMB and EMU 5σ galaxies $Cl^{gt}$ with WMAP 9 year's cosmological parameters. We used redshift between 0 and 1. Bottom: Normalized $Cl^{gt}$.

Figure 6 shows the cross-correlation power spectrum $Cl^{gt}$ with respect to l values. We can see from the normalized $Cl^{gt}$ that peak signal is achieved at multipole values around 20, which is also similar to the ideal (no shot noise and uniform distribution) survey discussed by [1]. Figure 7 provides $Cl^{gt}(\theta)$ estimates with covariance for the model we used.

To obtain galaxy auto-correlation for eq. 18 we have [34]:

$$Cl^{gg} = 4\pi \int_{kmin}^{kmax} \frac{dk}{k} \Delta^2(k)\{Wl^g(k)\}^2 \qquad (20)$$

We use the$Cl^{gg}$ approximation in [3] for large multipole(l) values. Cosmic variance for $Cl^{gg}$in this case is used as:

$$\Delta\, Cl^{gg} = \sqrt{\frac{2}{(2l+1)fsky}}\left(Cl^{gg} + \frac{1}{Ns}\right) \qquad (21)$$

Here fsky is taken as 0.5 or $\Delta\Omega = 2\pi$ showing final sky coverage after galactic foreground removal from both CMB and galaxy maps.

Signal to noise ratio of ISW-galaxy cross-correlation is dependent on maximum redshift and multipole ranges. To calculate signal to noise ratios we use:

$$\left(\frac{S}{N}\right)^2 = \sum_l \left(\frac{Cl^{gt}}{\Delta Cl^{gt}}\right)^2 \qquad (22)$$

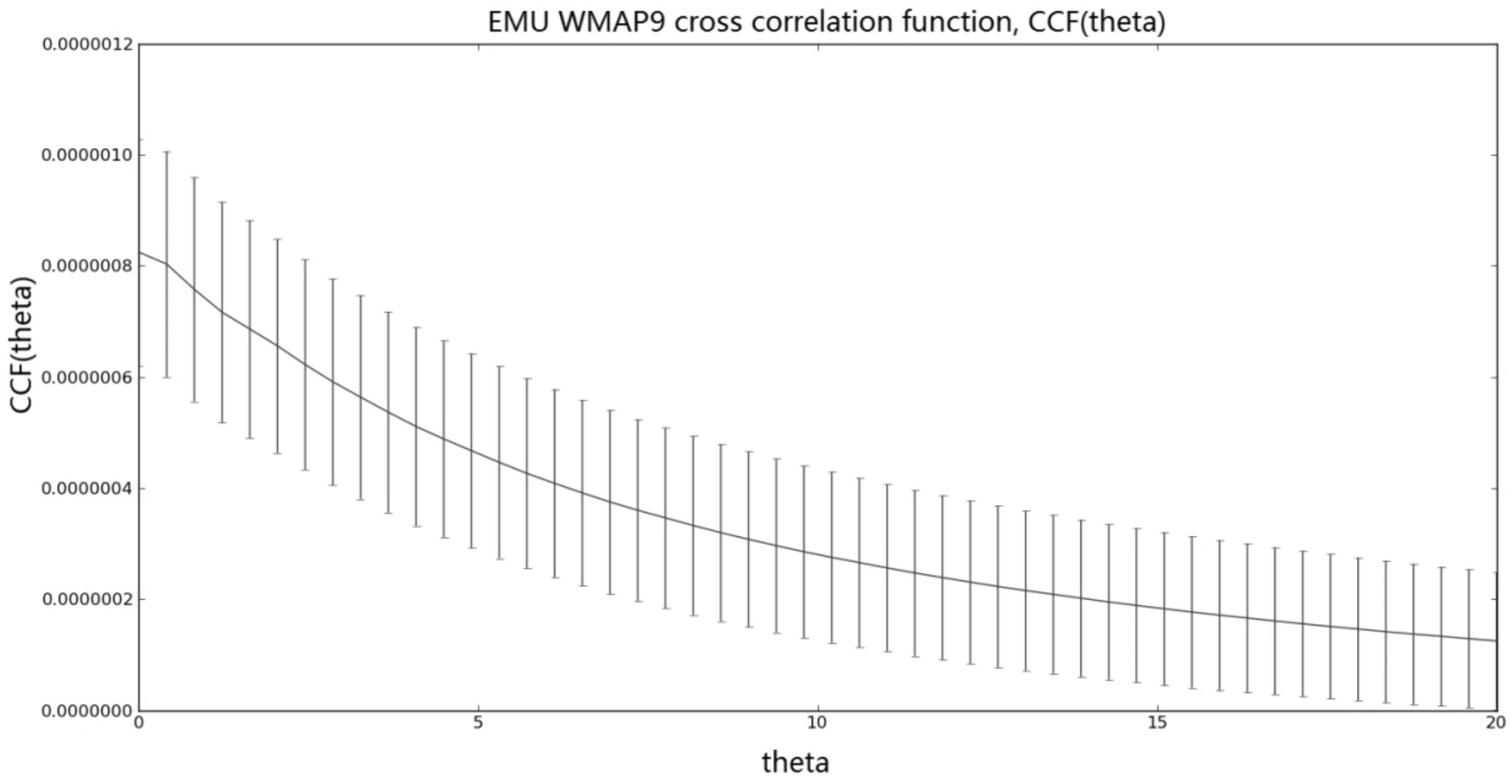

Figure 7-CCF(θ) for EMU-CMB angular cross correlation with WMAP 9 years cosmological parameters. We used 3 ≤ ‘l’ ≤ 100 for these calculations and z from 0 to 1.

Results in figure 7 are obtained using equations 17 and 19 which we can see are dependent on the maximum multipole (l) or lmax range. Figures 9 and 10, provide a clearer comparison between different lmax ranges.

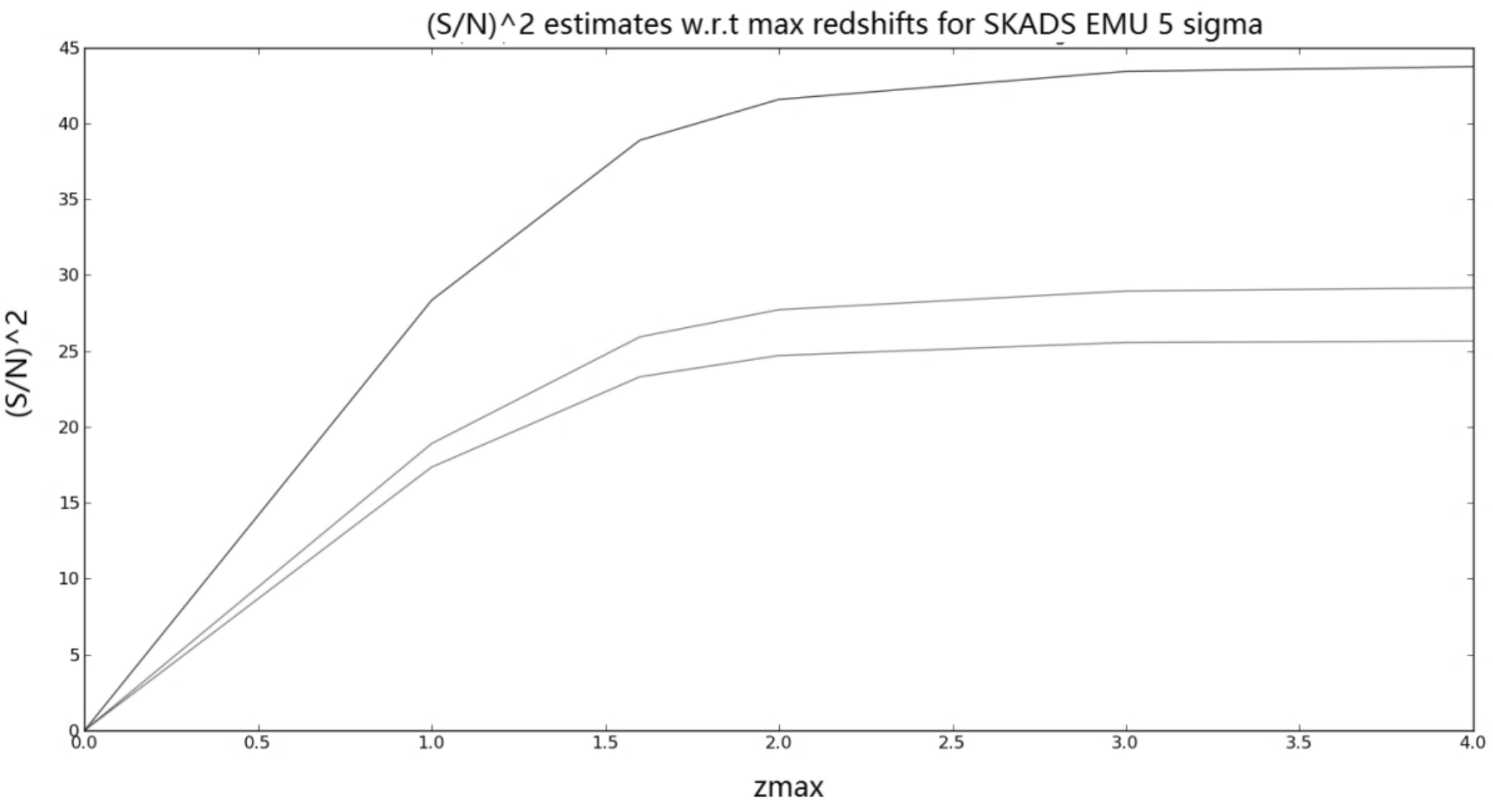

Figure 8-(S/N )^2 for different zmax ranges with fsky=0.75 (top curve), fsky=0.5 (middle) and fsky=0.5 with shot noise (bottom) .

We repeat the process for different zmax ranges and obtain a plot to see how S/N ratio works with different zmax ranges. In figure 8, top curve shows (S/N )^2 with fsky=0.75 (The original EMU-ASKAP coverage) without considering shot noise (1/Ns), middle curve shows (S/N )^2 with fsky=0.5 (after masking bright sources and other unwanted areas) without considering shot noise and bottom curve shows (S/N )^2 with fsky=0.5 after considering 1/Ns. We can see the noticeable improvement between zmax=1 and zmax=2 with SNR for fsky=.75 and without shot noise consideration from ≈5.3 for zmax=1 to ≈6.4 for zmax=2. After zmax=2, we do not see any significant improvement in (S/N)^2 results. Apart from other factors like galaxy bias, this can also be related to the source count for redshift bins which we can see in figure 5 where after z≈2, we have relatively very low number of sources per redshift bin. This also translates into relatively less increments in source count per steradian as we move after z≈2 even for very sensitive survey like EMU-ASKAP. The concentration of radio sources in redshift bins between 0.2 to 0.8 is similar to previous studies like [8].We can see in figure 3, the diminishing effect of increasing zmax range after zmax≈2 in reducing shot noise. This also shows the importance of ISW-signal in regions between redshifts0 and ≈ 1.5 [1]due to the dominance of dark energy which drives the accelerated expansion of the universe and plays a key role in late time ISW effect.

We can see in figure 10, some significant improvement in (S/N )^2 from lmax=50 to 100 but after that we see the effect is quickly diminishing. This also justifies the usual choice of lmax ∼ 100 in similar studies for ISW effect detection because almost all the signal is present in this region [1].

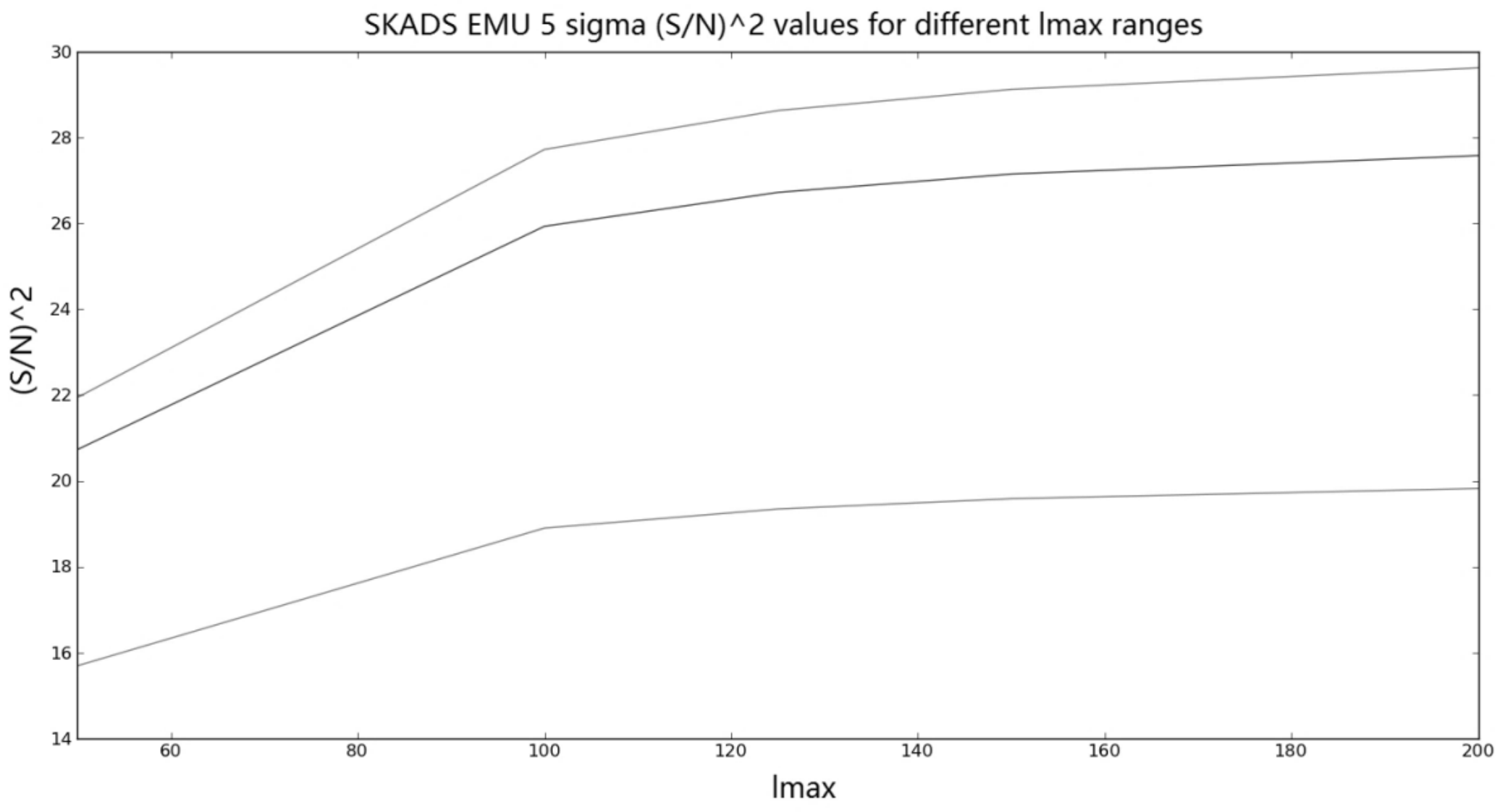

Figure 9-(S/N )^2 plot w. r.t different lmax ranges and fsky=0.5. Bottom curve shows zmax=1, middle shows zmax=1.6 and top shows zmax=2. Here shot noise was ignored.

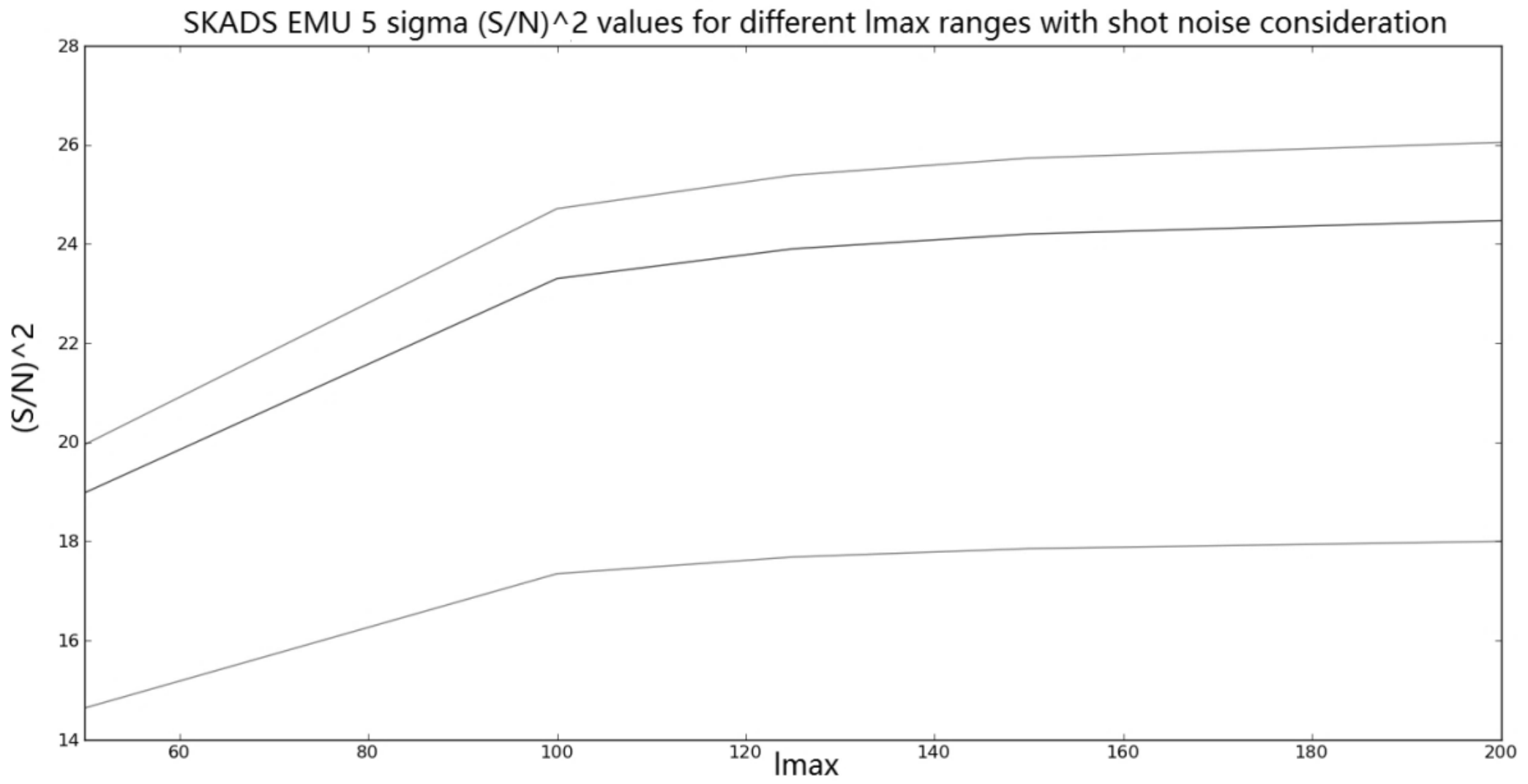

Figure 10-(S/N )^2 plot w.r.t different lmax ranges and fsky=0.5. Bottom shows zmax=1, middle shows zmax=1.6 and top shows zmax=2. Here shot noise was included in calculating (S/N )^2.

In an ideal survey with Ns → ∞, shot noise will not play much role, figures 1 and 2 show how more sensitive surveys can result in higher source count and so less shot noise, but again the issues related to resolution and confusion will limit our ability to achieve an ideal survey and shot noise will always play its part.

## 7. CONSTRAINTS ON COSMOLOGICAL PARAMETERS (Ωb and ΩΛ)

With zmax=1, we obtained constraints for ΩΛ and Ωb. Here Ωb is the baryonic matter density parameter.

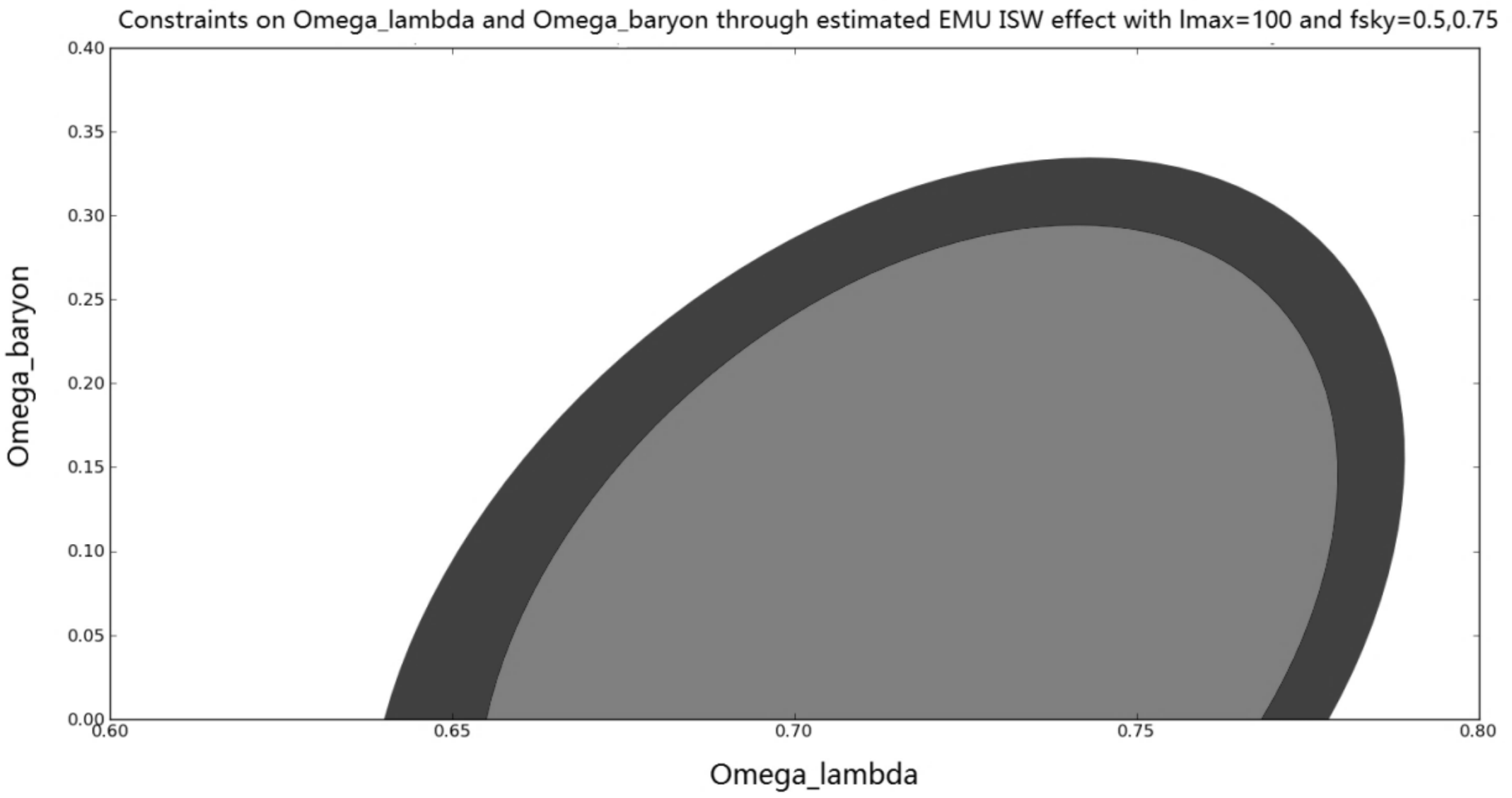

Figure 11-Inner ellipse shows constraints with fsky=0.75 and outer shows with fsky=0.5. Here we used lmax=100 and zmax=1.

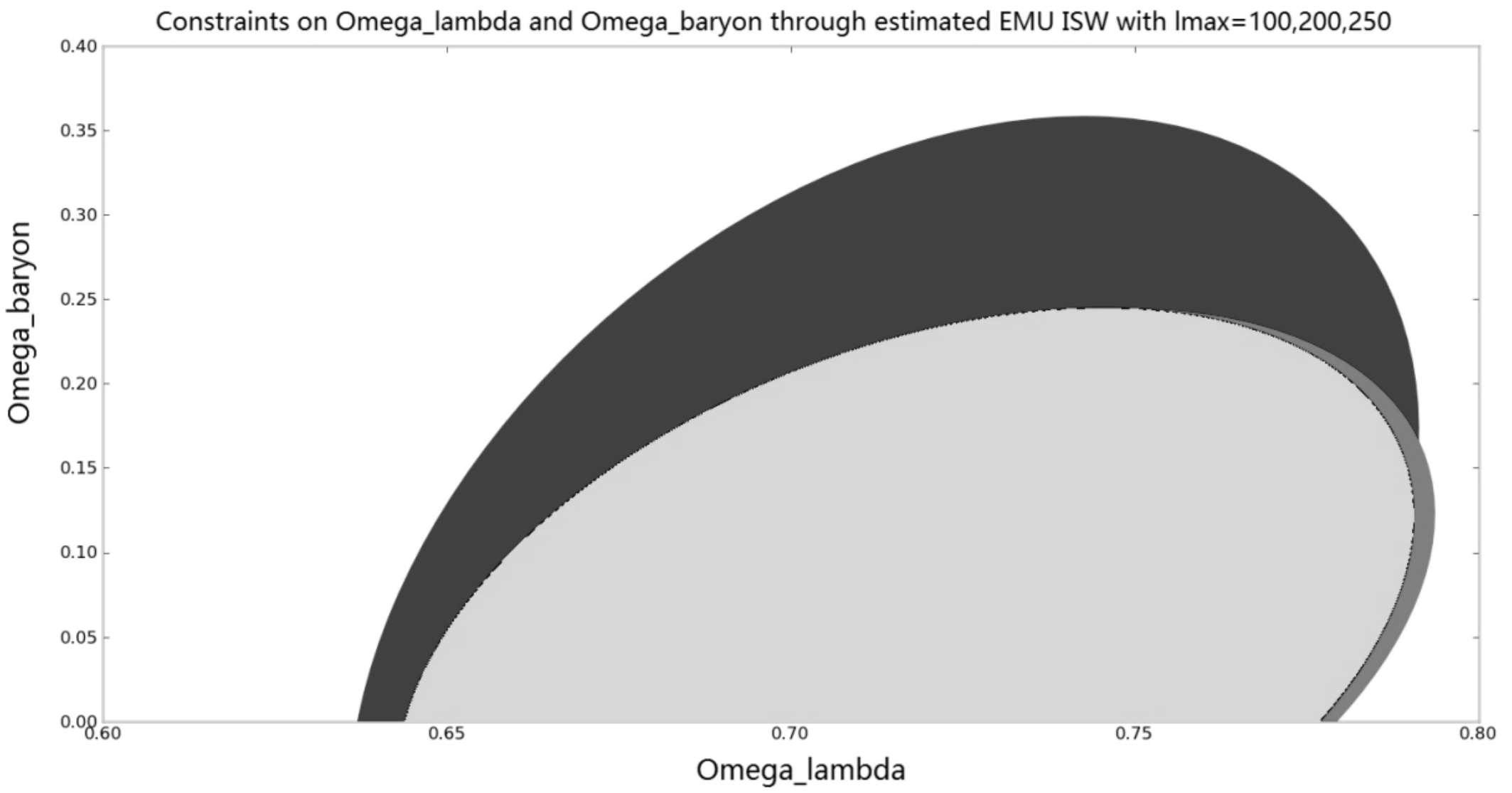

Figure 12-Here we have results with fsky=0.5 and zmax=1. Outer-most represents lmax=100, middle ellipse represents lmax=200 and inner-most represents lmax=250.

To obtain constraints ellipse, fisher matrix (F) is calculated as:

$$F^{ij}=fsky\sum_{l}\frac{(2l+1)\partial Cl^{gt}}{\partial\Theta i}\; Cov^{-1}(l)\,\frac{\partial Cl^{gt}}{\partial\Theta j} \qquad (23)$$

Where,

$Cov^{-1}=[Cl^{gt}]^2+(Cl^{ISW}+Nl^{cmb})(Cl^{gg} + Nl^{gg})$

We have, $Cl^{ISW}+Nl^{cmb}\approx Cl^{tt}$ or Cl of CMB as we correlate entire CMB signal or Cltt and not just ISW signal. $Nl^{gg}$ is mainly shot noise and so can be written as 1/Ns. Θ represents set of cosmological parameters being constrained.

In figure 11, we can see the effect of sky coverage on constraints ellipse. EMU-ASKAP will essentially cover around 75 % of the sky but due to foreground masking, we will end up with an fsky ≈ 0.5. However, a similar northern sky survey, WODAN [36] will compliment EMU-ASKAP data and will increase the overall sky coverage to 75% again to measure the ISW effect.

Figure 10 shows the diminishing impact of increasing lmax range after lmax=100 over SNR. In figure 12, we can see the effect of increasing lmax range and how the parameter constraints are tightened with increasing lmax ranges after lmax=100. For a reliable analysis, we need to measure the effect over various ranges of redshift, lmax and sky coverage.

## 8. CONCLUSION

In this study, we estimated ISW signal from EMU-ASKAP 5σ survey using WMAP9 parameters in the Lambda CDM model. We also calculated rms confusion limit and how this can help in estimating rms position uncertainty for the planned survey. We have also performed error analysis and saw how things like shot noise can influence survey data results and analysis of ISW effect. We also presented signal to noise ratio analysis for different zmax ranges and saw some significant improvements prior to zmax ≈ 2 range. We also constrained ΩΛ and Ωb using EMU-WMAP9 cross correlation with zmax=1 and lmax=100 and observed improvements to constraints with increase in lmax ranges and sky coverage. In the end, we can say that we need to test the data for a range of redshift and multipole ranges. We also need to take care of observational issues like shot noise, confusion accuracy and position accuracy, to get reliable results and to avoid any confusion related to the significance of the results.